# Hodge Duality and Cosmological Constant

Hitoshi Nishino[1] and Subhash Rajpoot[2]

*Department of Physics & Astronomy*
*California State University*
*1250 Bellflower Boulevard*
*Long Beach, CA 90840*

## Abstract

We present a simple mechanism to eliminate cosmological constants both in supersymmetric and non-supersymmetric theories. This mechanism is based on the Hodge (Poincaré) duality between a 0-form and $D$-form field strengths in $D$-dimensional space-time. The new key ingredient is the introduction of an extra Chern-Simons term into the $D$-form field strength $H$ dual to the 0-form field strength. Our formulation can be also made consistent with supersymmetry. Typical applications to four-dimensional $N = 1$ supergravity and to ten-dimensional type IIA supergravity are given. The success of our formulation for both supersymmetric and non-supersymmetric systems strongly indicates the validity of our mechanism even after supersymmetry breakings at the classical level. Our mechanism may well be applicable to quantized systems, at least for supersymmetric cases with fundamental D-brane actions available.

---

[1]E-Mail: hnishino@csulb.edu

[2]E-Mail: rajpoot@csulb.edu

## 1. Introduction

The smallness or zero-ness of the cosmological constant to the accuracy of $10^{-120}$ in the dimension of $(\text{mass})^4$ has been mystifying theoretical particle physicists since its inception [1]. The problem is not whether we *can* have a model with a zero or extremely small cosmological constant, but that we *must* always perform 'artificial' fine-tunings with the accuracy of $10^{-120}$ to remove the cosmological constant both at classical and quantum levels. To date, there seems to be no satisfactory mechanism for its explanation, either in fundamental high energy theories including superstring theory [2], M-theory [3], or in cosmological models [1]. Supergravity theory with local supersymmetry [4] was once expected to provide a nice mechanism of fine-tuning the cosmological constant to be exactly zero. However, it turned out that supergravity theory also needs an artificial fine-tuning [5][6], even though the quantum divergences in such theories are better controlled by supersymmetry. The dilaton scale invariance common in supergravity originated from string tree level invariance [7] was also expected to play an important role. But the questions are whether such a tree-level symmetry is maintained at higher-loops, and how to avoid a massless dilaton [2]. Some attempts have also been made in quantum cosmology, based on baby universes or wormholes [8], but there seems to be the difficulty of forbidding the creation of undesirably 'big' baby universes [9].

Ever since Einstein's blunder [10], the cosmological constant has been an enigma in general relativity. Lagrangian formulation of general relativity admits it, while no known symmetry forbids it, and up until recently, it was not required empirically. Recent Type Ia supernova observation [11] provides evidence that the universe is accelerating at a greater rate now than in the past, and implies a non-zero cosmological constant $\Lambda \neq 0$. An interesting implication of this is that energy density $\Omega_\Lambda$ associated with non-zero $\Lambda$ is of the same order of magnitude as the matter density of the universe, giving rise to the so-called second cosmological problem. This has led to a flurry of activity explaining the two cosmological problems, and involves the anthropic principle [1][12], quintessence [13], new interactions, extra dimensions, phase transitions, and space-time fluctuations. However, more data are required before definite conclusions can be drawn. In our present work, we will assume $\Lambda = 0$, and present a dynamical mechanism that uniquely determines the value $\Lambda = 0$.

Understanding the zero-ness or smallness of the cosmological constant has been also attempted in terms of 'unimodular gravity' formulation [10][14] or its supersymmetric version 'unimodular supergravity' [15]. In these formulations, the zero-ness or smallness of the cosmological constant $\Lambda$ is less serious, because the value of $\Lambda$ is no longer something put in by hand, but is regarded as an 'initial condition'. However, we still have to input such a fine-tuned initial condition at the beginning, based on a special constraint $\det(e_\mu{}^m) = 1$ which seems rather *ad hoc*, unfounded or artificial.

One important development in the 1990's is the discovery of dualities playing crucial roles in string physics, such as S-, T- and U-dualities [16]. Dualities are similar to symmetries, but they have distinct features, such as describing different phases of coupling constants. It is a general conviction that the cosmological constant problem may be solved by invoking symmetries or something similar. Therefore it is a natural step to consider certain dualities as a trial solution to this problem.

For example, a scenario of supersymmetry breaking maintaining the zero cosmological constant in three-dimensions (3D) with conical singularities and strong/weak duality [17]. was proposed by Witten [18]. Such a conical space-time has a deficit angle, lifting the degeneracy between bosons and fermions. An explicit model in 3D has been presented based on 3D, $N = 2$ supergravity, and mechanisms applicable to 4D based on strong/weak coupling duality have been also proposed [18][19]. However, it is not clear how such duality with 3D can help for building realistic models with required Lorentz invariance in 4D.

The duality we rely on in our present paper is the so-called Hodge (Poincaré) duality, *i.e.*, the relationship between the $r$-form and $(D-r)$-form field strengths in $D$ space-time dimensions. Typical examples are $\partial_\mu \varphi = (1/3!)e^{-1}\epsilon_\mu{}^{\nu\rho\sigma}G_{\nu\rho\sigma}$ between the 1-form $d\varphi$ and the 3-form $G \equiv 3dB$ field strengths in 4D, or 3-form vs. 7-form field strengths in 10D [20].

The Hodge (Poincaré) duality used in this paper is between a 0-form and a $D$-form in $D$ space-time dimensions. The '0-form field strength' is understood better in mathematical terms: Let $\alpha$ be a '0-form field strength'. Since it is a field strength, it must satisfy a 'Bianchi identity' $d\alpha = 0$. As $\alpha$ is a 0-form, the scalar field $\alpha$ is nothing but a constant. Even though this fact itself seems to have poor content, its significance will be gradually elucidated in this paper. Let the cosmological constant term in a lagrangian in $D$-dimensions be

$$\sqrt{-g}\,\Lambda \equiv \tfrac{1}{2}a\,m^2\,e \quad , \tag{1.1}$$

where $a$ is a non-zero numerical constant. We emphasize here that $\Lambda$ can contains not only 'classical' cosmological constant, but also 'quantum' cosmological constant. We use this notation (1.1), in order to follow the common usage in supergravity theories in diverse dimensions [21]. We next regard $m$ as such a '0-form field strength'. Now we invoke the duality relationship between a 0-form $m$ and a maximal-rank $D$-form field strength $H$ to be

$$m = \frac{1}{D!}\,e^{-1}\epsilon^{\mu_1\cdots\mu_D}H_{\mu_1\cdots\mu_D} \quad , \tag{1.2}$$

where $H = DdC+$ 'Chern-Simons', where $C$ is a $(D-1)$-form potential. Even though these points are similar to the duality transformation dictated [22], there is an essential difference. In our system, the existence of the Chern-Simons term in the field strength $H$ plays an important role, in addition to the conventional duality transformation [22].

Introducing maximal-rank field strengths is not a new idea. For example, the compactifications [23] of 11D supergravity [24] or $N=8$ supergravity in 4D [25] have motivated the introduction of 4-form field strength $F_{\mu\nu\rho\sigma}$ [26]. However, our formulation differs from these previous ones by the additional 'Chern-Simons term' within the maximal-rank field strength $H$.

In this paper, we will show that maintaining this duality with the $H$-field strength including the extra Chern-Simons term at the lagrangian level excludes the possible non-zero cosmological constant at the field equation level. This mechanism works for both supersymmetric and non-supersymmetric systems. As interesting applications, we present as examples $N=1$ supergravity in 4D [5] and massive type IIA supergravity in 10D [27][28][29].

## 2. General Mechanism in Arbitrary $D$-Dimensions

Consider a general $D$-dimensional space-time, and let $\mathcal{L}_0$ be any lagrangian with the Hilbert action term: $-(1/4)eR$. Separate the cosmological constant term $e\Lambda$ from $\mathcal{L}_0$, as[3]

$$\mathcal{L}_1 \equiv \mathcal{L}_0 + e\Lambda \equiv \mathcal{L}_0 + \tfrac{1}{2}am^2 e \ , \tag{2.1}$$

where $am^2/2 \equiv \Lambda$ as in (1.1). We now perform the duality transformation from the zero-form $m$ to the $D$-form field strength $H$. To this end, we first rewrite $m$ as the scalar field $M(x)$, and then add the constraint lagrangian dictating the constancy of $M$ [22]:

$$\mathcal{L}_{\text{tot}} \equiv \mathcal{L}_0 + \tfrac{1}{2}aeM^2 + \mathcal{L}_c \ , \tag{2.2a}$$

$$\mathcal{L}_c \equiv -\tfrac{1}{D!}\, a\, \epsilon^{\mu_1\cdots\mu_D}\, M\, H_{\mu_1\cdots\mu_D} \ , \tag{2.2b}$$

$$H_{\mu_1\cdots\mu_D} \equiv D\, \partial_{[\mu_1} C_{\mu_2\cdots\mu_D]} + M\, K_{\mu_1\cdots\mu_D} \ . \tag{2.2c}$$

Compared with [22], the last term in (2.2c) is a 'generalized' Chern-Simons term. The $K$ is a $D$-form potential field whose $(D+1)$-form field strength is 'formally' defined, but it does not actually exist in $x$-space-time, due to its over-rank $D+1 > D$. The $MK$-term in (2.2c) is a generalized Chern-Simons term for the following reasons. First, it is a product of a '0-form' field strength $M$ and a $D$-form potential $K$. Second, the exterior derivative of (2.2c) formally gives a product of a 0-form and a $(D+1)$-form 'field strengths':

$$(D+1)dH = ML \ , \tag{2.3}$$

where $L \equiv (D+1)dK$ is the $(D+1)$-form field strength: $L_{\mu_1\cdots\mu_{D+1}} \equiv (D+1)\partial_{[\mu_1}K_{\mu_2\cdots\mu_{D+1}]}$.

The Chern-Simons term in (2.2c) is motivated by the superspace formulation [29] of massive Type IIA supergravity [28][27]. In this sense, the Chern-Simons term in (2.2c) is not artificially put in by hand, but has been well-motivated in terms of local supersymmetry in superspace [28][29].

---
[3]We use the metric $(\eta_{mn}) = \text{diag.}\,(+,-,-,\cdots,-)$ in this paper.

The $C$-field equation is now

$$\partial_\mu M = 0 \implies M = \text{const.} \, , \tag{2.4}$$

while the $M$-field equation yields[4]

$$M \doteq + \tfrac{1}{D!} e^{-1} \epsilon^{[D]} H_{[D]} + \tfrac{1}{D!} e^{-1} \epsilon^{[D]} M K_{[D]} \, . \tag{2.5}$$

Our new feature is that the $K$-field equation yields the zero cosmological constant

$$M \doteq 0 \implies \Lambda \doteq 0 \, . \tag{2.6}$$

Therefore, the last term in (2.5) also vanishes, while yielding the desirable duality $M \doteq + (1/D!) e^{-1} \epsilon^{[D]} H_{[D]}$.

We are using the Hodge duality with the extra Chern-Simons term $MK$ in $H$ as the important guiding principle. There might be many other mechanisms yielding the same effect to remove the cosmological constant. For example, the $C$-field in (2.2c) can be absorbed into the field redefinition of the $K$-field. By the same token, a lagrangian term $am\widetilde{C}$ with a scalar density $\widetilde{K}$ dual to $K$ can set $m = 0$. Furthermore, an extreme method is just to put $\Lambda = 0$ 'by hand' from the outset. However, these mechanisms are not well founded by any principles, geometrical formulations, or dynamical field equation to yield the desired results. In this context, we re-emphasize the importance of 0-form $D$-form Hodge duality with the extra Chern-Simons term $MK$ in the field strength $H$.

As is easily seen, our mechanism is based on a very general form of cosmological constant in (2.1). Therefore, our formulation is also applicable a system with spontaneous symmetry breakings, such as $SU(2)_\text{W} \times U(1)_Y \to U(1)_\text{em}$. Moreover, our mechanism may well be compatible with quantized systems, as long as the cosmological constant $\Lambda$ includes also quantum effects with loop corrections. We will be back to this point shortly.

## 3. Application to 4D Supergravity

As the most important application of our mechanism, we use it for $N = 1$ supergravity in 4D. For simplicity, we adopt the formulation of massive gravitino with a cosmological constant [5], with no matter multiplet. We start with the lagrangian

$$\mathcal{L}_\text{4D} = -\tfrac{1}{4} e R - \tfrac{i}{2} e (\overline{\psi}_\mu \gamma^{\mu\nu\rho} D_\nu \psi_\rho) - \tfrac{1}{2} M (\overline{\psi}_\mu \gamma^{\mu\nu} \psi_\nu) + \tfrac{3}{2} e M^2 - \tfrac{1}{8} \epsilon^{\mu\nu\rho\sigma} M H_{\mu\nu\rho\sigma} \, , \tag{3.1}$$

---

[4]The index $[n]$ stands for totally antisymmetrized $n$ indices in order to save space, e.g., $\epsilon^{[D]} K_{[D]} \equiv \epsilon^{\mu_1 \cdots \mu_D} K_{\mu_1 \cdots \mu_D}$.

where for $D=4$ and $a=3$ in (2.2):

$$H_{\mu\nu\rho\sigma} \equiv 4\partial_{[\mu}C_{\nu\rho\sigma]} + MK_{\mu\nu\rho\sigma} \ . \tag{3.2}$$

All the terms in (3.1) except the last one are exactly the same as the original $N=1$ supergravity lagrangian with a cosmological constant [5] with the mass parameter $m$ replaced by $M$.

The action $I_{\rm 4D} \equiv \int d^4x \, \mathcal{L}_{\rm 4D}$ is invariant under $N=1$ supersymmetry

$$\delta_Q e_\mu{}^m = -i(\overline{\epsilon}\gamma^m\psi_\mu) \ , \quad \delta_Q \psi_\mu = D_\mu\epsilon + \tfrac{i}{2}M\gamma_\mu\epsilon \ ,$$
$$\delta_Q C_{\mu\nu\rho} = +i(\overline{\epsilon}\gamma_5\gamma_{[\mu\nu}\psi_{\rho]}) \ , \quad \delta_Q K_{\mu\nu\rho\sigma} = 0 \ , \quad \delta_Q M = 0 \ . \tag{3.3}$$

The transformation $\delta_Q C$ is rather easily fixed, in such a way that all the $\partial M$-dependent terms are cancelled by $\delta_Q C$ *via* the $MH$-term. The fact that the $M$-field is invariant under supersymmetry poses no problem, due to its constancy. Similarly, as will be seen, all the effect of the $K$-field disappears from the whole field equations justifies the invariance of $K$ under supersymmetry.

We can verify the closure of supersymmetry, as in the conventional system. In particular, the closure on the $C$-field is easily verified, up to a local gauge transformation $\delta_\lambda C_{\mu\nu\rho\sigma} = 4\partial_{[\mu}\lambda_{\nu\rho\sigma]}$ [22][28].

The new feature of our system compared with [22] is the existence of the extra Chern-Simons term with the $K$-field. In fact, the $K$-field equation yields

$$M \doteq 0 \quad \Longrightarrow \quad \Lambda \doteq 0 \ . \tag{3.4}$$

Once this is satisfied, the $C$-field equation

$$M \doteq \text{const.} \tag{3.5}$$

is automatically satisfied. Accordingly, the $M$-field equation is simply

$$M \doteq +\tfrac{1}{24}e^{-1}\epsilon^{\mu\nu\rho\sigma}\widehat{H}_{\mu\nu\rho\sigma} \ , \tag{3.5}$$

where $\widehat{H}$ is the usual supercovariantized field strength [4]. All the fermion bilinear terms in (3.4) are completely absorbed into the supercovariantization of the field strength $H$, which also reconfirms the transformation rule $\delta_Q C$. Note that we skipped the $MK$-term like that in (2.5), because of the vanishing of the $M$-field itself. Needless to say, due to the vanishing $M$-field, each side of (3.5) vanish on-shell. The on-shell duality relationship (3.5) is also consistent with supersymmetry, just as in the usual duality transformation [22].

If we do not include the Chern-Simons term $MK$ in $H$, and apply the usual duality transformation in [22] to the supergravity system in 4D [5], the scalar field $M$ in the cosmological constant term $3eM^2/2$ can take any arbitrary value, as in [22][28]. In our formulation, however, the dynamical $K$-field equation *uniquely* fixes $\Lambda \doteq 0$.

We have thus seen that our mechanism is applicable to supergravity in 4D, consistently with local supersymmetry. Applications to supergravity with more matter multiplets, even with supersymmetry breakings *via* super Higgs effects [6] can be accomplished without further essential problems [30].

## 4. Application to 10D Massive Type IIA Supergravity

As another interesting application, we look into the case of massive type IIA supergravity in 10D [27] which is obtained from massless type IIA supergravity [31]. A cosmological constant term in Massive type IIA supergravity [27] is actually identified with a dilaton potential term proportional to $m^2 e e^{-5\varphi}$. We show that our mechanism constrains the mass parameter $m$ to be zero, uniquely yielding a zero cosmological constant. Most importantly, this can be done consistently with local $N=(1,1)$ supersymmetry in 10D.

One caveat is that we need the massive Type IIA supergravity lagrangian with the smooth limit of $m \to 0$. In fact, the lagrangian in [27] does not do the job, because the field strength $F_{\mu\nu}$ has been absorbed into the $B_{\mu\nu}$-field, and the limit $m \to 0$ is not smooth. For this reason, we need to reconstruct the original lagrangian where the field strength $F_{\mu\nu}$ is present.

Keeping this caveat in mind, we apply our mechanism to Type IIA supergravity in 10D [31][27]. The field content of our system is $(e_\mu{}^m, \psi_\mu, A_{\mu\nu\rho}, B_{\mu\nu}, A_\mu, \chi, \varphi, M, C_{\mu_1\cdots\mu_9}, K_{\mu_1\cdots\mu_{10}})$, where only $M$, $C$ and $K$ are our new fields. Our lagrangian is now fixed to be

$$\begin{aligned}
\mathcal{L}_{10D} =\ & -\tfrac{1}{4}eR - \tfrac{i}{2}e(\bar\psi_\mu \gamma^{\mu\nu\rho} D_\nu \psi_\rho) + \tfrac{i}{2}e(\overline\chi \gamma^\mu D_\mu \chi) + \tfrac{1}{2}e(\partial_\mu\varphi)^2 - \tfrac{1}{48}ee^{-\varphi}(F'_{[4]})^2 + \tfrac{1}{12}ee^{2\varphi}(G_{[3]})^2 \\
& - \tfrac{1}{4}ee^{-3\varphi}(F'_{\mu\nu})^2 + \tfrac{1}{96}ee^{-\varphi/2}\Big[(\overline\psi^\mu \gamma_{[\mu|}\gamma^{[4]}\gamma_{|\nu]}\psi^\nu) - \tfrac{i}{\sqrt{2}}(\overline\psi_\mu \gamma^{[4]}\gamma^\mu\chi) + \tfrac{3}{4}(\overline\chi\gamma^{[4]}\chi)\Big]F'_{[4]} \\
& + \tfrac{1}{24}ee^{\varphi}\Big[i(\overline\psi^\mu \gamma_{11}\gamma_{[\mu|}\gamma^{[3]}\gamma_{|\nu]}\psi^\nu) - \sqrt{2}(\overline\psi_\mu \gamma^{[3]}\gamma^\mu\chi)\Big]G_{[3]} - \tfrac{1}{\sqrt{2}}e(\overline\psi_\mu \gamma_{11}\gamma^\nu\gamma^\mu\chi)\partial_\nu\varphi \\
& + \tfrac{1}{8}ee^{-3\varphi/2}\Big[(\overline\psi^\mu \gamma_{11}\gamma_{[\mu|}\gamma^{\rho\sigma}\gamma_{|\nu]}\psi^\nu) - \tfrac{3i}{\sqrt{2}}(\overline\psi_\mu \gamma^{\rho\sigma}\gamma^\mu\chi) + \tfrac{5}{4}(\overline\chi\gamma_{11}\gamma^{\rho\sigma}\chi)\Big]F'_{\rho\sigma} \\
& + \tfrac{1}{1152}\epsilon^{\mu\nu\rho\sigma\tau\lambda\omega\psi\varphi\chi}\Big(F_{\mu\nu\rho\sigma}F_{\tau\lambda\omega\psi}B_{\varphi\chi} + 12F_{\mu\nu\rho\sigma}F_{\tau\lambda}B_{\omega\psi}B_{\varphi\chi} + 48F_{\mu\nu}F_{\rho\sigma}B_{\tau\lambda}B_{\omega\psi}B_{\varphi\chi} \\
& \qquad + 4MF_{\mu\nu\rho\sigma}B_{\lambda\tau}B_{\omega\psi}B_{\varphi\chi} + 36MF_{\mu\nu}B_{\rho\sigma}B_{\tau\lambda}B_{\omega\psi}B_{\varphi\chi} + \tfrac{36}{5}M^2 B_{\mu\nu}B_{\rho\sigma}B_{\tau\lambda}B_{\omega\psi}B_{\varphi\chi}\Big) \\
& + \tfrac{1}{8}eMe^{-5\varphi/2}(\overline\psi_\mu \gamma^{\mu\nu}\psi_\nu) - \tfrac{5i}{8\sqrt{2}}eMe^{-5\varphi/2}(\overline\psi_\mu\gamma_{11}\gamma^\mu\chi) - \tfrac{21}{32}eMe^{-5\varphi/2}(\overline\chi\chi) \\
& - \tfrac{1}{8}ee^{-5\varphi}M^2 + \tfrac{1}{4(10!)}\epsilon^{[10]}MH_{[10]} \ . \tag{4.1}
\end{aligned}$$

Our $\gamma_{11}$ satisfies $(\gamma_{11})^2 = +I$. The conventional field strengths are defined by [27][28]

$$F'_{\mu\nu} \equiv F_{\mu\nu} + MB_{\mu\nu} \ , \quad F_{\mu\nu} \equiv 2\partial_{[\mu}A_{\nu]} \ , \quad G_{\mu\nu\rho} \equiv 3\partial_{[\mu}B_{\nu\rho]} \ ,$$

$$F'_{\mu\nu\rho\sigma} \equiv F_{\mu\nu\rho\sigma} + 12B_{[\mu\nu}F'_{\rho\sigma]} - 6MB_{[\mu\nu}B_{\rho\sigma]} \ , \quad F_{\mu\nu\rho\sigma} \equiv 4\partial_{[\mu}A_{\nu\rho\sigma]} \ , \tag{4.2}$$

while $a = -1/4$ in (2.2), and our $H_{[10]}$ contains Chern-Simons $MK$-term as in (2.2) for $D = 10$:

$$H_{\mu_1\cdots\mu_{10}} \equiv 10\partial_{[\mu_1} C_{\mu_2\cdots\mu_{10}]} + MK_{\mu_1\cdots\mu_{10}} \ . \tag{4.3}$$

Our action $I_{10D} \equiv \int d^{10}x \, \mathcal{L}_{10D}$ is invariant under supersymmetry

$$\delta_Q e_\mu{}^m = -i(\bar\epsilon\gamma^m\psi_\mu) \ , \quad \delta_Q\varphi = -\tfrac{1}{\sqrt{2}}(\bar\epsilon\gamma_{11}\chi) \ , \tag{4.4a}$$

$$\delta_Q\psi_\mu = +D_\mu(\widehat\omega)\epsilon - \tfrac{i}{32}e^{-3\varphi/2}\gamma_{11}(\gamma_\mu{}^{\rho\sigma} - 14\delta_\mu{}^\rho\gamma^\sigma)\epsilon F'_{\rho\sigma}$$
$$- \tfrac{1}{48}e^{\varphi}\gamma_{11}(\gamma_\mu{}^{\nu\rho\sigma} - 9\delta_\mu{}^\nu\gamma^{\rho\sigma})\epsilon G_{\nu\rho\sigma}$$
$$+ \tfrac{i}{128}e^{-\varphi/2}(\gamma_\mu{}^{\nu[3]} - \tfrac{20}{3}\delta_\mu{}^\nu\gamma^{[3]})\epsilon F'_{\nu[3]} - \tfrac{i}{32}Me^{-5\varphi/2}\gamma_\mu\epsilon \ , \tag{4.4b}$$

$$\delta_Q A_{\mu\nu\rho} = +\tfrac{3}{2}e^{\varphi/2}(\bar\epsilon\gamma_{[\mu\nu}\psi_{\rho]}) + \tfrac{i}{4\sqrt{2}}e^{\varphi/2}(\bar\epsilon\gamma_{11}\gamma_{\mu\nu\rho}\chi) - 6B_{[\mu\nu|}(\delta_Q A_{|\rho]}) \ , \tag{4.4c}$$

$$\delta_Q A_\mu = +\tfrac{1}{2}e^{3\varphi/2}(\bar\epsilon\gamma_{11}\psi_\mu) + \tfrac{3i}{4\sqrt{2}}e^{3\varphi/2}(\bar\epsilon\gamma_\mu\chi) \ , \tag{4.4d}$$

$$\delta_Q B_{\mu\nu} = -ie^{-\varphi}(\bar\epsilon\gamma_{11}\gamma_{[\mu}\psi_{\nu]}) + \tfrac{1}{2\sqrt{2}}e^{-\varphi}(\bar\epsilon\gamma_{\mu\nu}\chi) \ , \tag{4.4e}$$

$$\delta_Q\chi = +\tfrac{i}{\sqrt{2}}\gamma_{11}\gamma^\mu\epsilon\partial_\mu\varphi + \tfrac{3}{8\sqrt{2}}e^{-3\varphi/2}\gamma^{\rho\sigma}\epsilon F'_{\rho\sigma}$$
$$+ \tfrac{i}{12\sqrt{2}}e^{\varphi}\gamma^{[3]}\epsilon G_{[3]} + \tfrac{1}{96\sqrt{2}}e^{-\varphi/2}\gamma_{11}\gamma^{[4]}\epsilon F'_{[4]} + \tfrac{5}{8\sqrt{2}}Me^{-5\varphi/2}\epsilon \ , \tag{4.4f}$$

$$\delta_Q C_{\mu_1\cdots\mu_9} = -9e^{-5\varphi/2}(\bar\epsilon\gamma_{11}\gamma_{[\mu_1\cdots\mu_8}\psi_{\mu_9]}) + \tfrac{5i}{2\sqrt{2}}e^{-5\varphi/2}(\bar\epsilon\gamma_{\mu_1\cdots\mu_9}\chi) \ , \tag{4.4g}$$

$$\delta_Q K_{\mu_1\cdots\mu_{10}} = 0 \ , \quad \delta_Q M = 0 \ , \tag{4.4h}$$

The transformation rule for $C$ is fixed, such that the new terms compared with [27] are invariant under supersymmetry up to a total divergence. Despite the presence of the dilaton exponents, our mechanism works smoothly also in this massive type IIA supergravity in 10D.

The closure of supersymmetry can be also easily verified, as in the conventional system [31][27]. In particular, the closure on the $C$-field is fixed in such a way that all the $\partial M$-dependent terms can be cancelled by $\delta_Q C$ [22][28].

The $K$-field equation yields

$$M \doteq 0 \implies \Lambda \doteq 0 \ , \tag{4.8}$$

which automatically satisfies the $C$-field equation

$$M \doteq \text{const.} \tag{4.9}$$

which is automatically satisfied by (4.8). The $M$-field equation yields the desirable duality relationship

$$M \doteq \tfrac{1}{10!} e^{-1} e^{5\varphi} \epsilon^{[10]} \widehat{H}_{[10]} - \tfrac{21}{8} e^{5\varphi/2} (\overline{\chi}\chi) \ , \tag{4.10}$$

which is also invariant under supersymmetry. Note the presence of the $(\overline{\chi}\chi)$-term in the duality relationship (4.10) that can be traced back to the involvement of the exponential function $e^{5\varphi}$ with $\widehat{H}$, which generates the $\chi H$-term, necessitating the $(\overline{\chi}\chi)$-term. Our mechanism results in a system of type IIA supergravity with zero dilaton potential or zero cosmological constant.

Note that the consequence $M \doteq 0$ is not artificially put in by hand, but required by the $K$-field equation. Out of an innumerable number of values of $M$ keeping the supersymmetric invariance of the action, the $K$-field equation picks up the unique value $M \doteq 0$. In this sense, the vanishing of the cosmological constant is required by the dynamics of our system, but not artificially installed by hand.

Even though the basic mechanism is parallel to the previous 4D case, the 10D case has slight differences. First, we have regarded the dilaton potential term as the 'cosmological constant' term. Second, we need the special dilaton dependence in exponents, such as $e^{5\varphi}$ in the duality (4.7). while this slightly complicates the system, its presence does not affect the essential structure of our mechanism.

The success of our mechanism applied to type IIA supergravity in 10D [31][27] is not a coincidence, but in a sense is a natural result. This is because the original introduction of the maximal-rank field strength [28] was motivated by the study of massive type IIA supergravity [27] which was further re-formulated in superspace [29]. Note also that our present formulation here differs from that in [28] due to the generalized Chern-Simons term $MK$ introduced into the maximal field strength $H$ which was motivated by superspace formulation [29].

The 10-th rank field $K$ is not *ad hoc* or artificial, but is clarified by the super ninebrane action in 10D formulated on 10D super world-volume [29]:[5]

$$I_{p=9} \equiv \int d^{10}\sigma \left[ +\tfrac{1}{2}\sqrt{-g}\, g^{ij} \Pi_i{}^a \Pi_{ja} - 4\sqrt{-g} + \tfrac{1}{10!} \epsilon^{i_1 \cdots i_{10}} \Pi_{i_1}{}^{A_1} \cdots \Pi_{i_{10}}{}^{A_{10}} K_{A_{10} \cdots A_1} \right] \ , \tag{4.11}$$

where $K_{A_1 \cdots A_{10}}$ is the 10-th rank potential superfield, while $i, j, \cdots = 0, 1, \cdots, 9$ are the coordinates for the 10D super world-volume. The set of constraints to be used is so-called '$\beta$-function favored constraints' (BFFC) to simplify the action above [29]. This action is invariant under the fermionic $\kappa$-symmetry:

$$\delta_\kappa E^{\underline{\alpha}} = \tfrac{1}{2}(I+\Gamma)^{\underline{\alpha}}{}_{\underline{\beta}} \kappa^{\underline{\beta}} \ , \quad \delta_\kappa E^a = 0 \ ,$$
$$\Gamma_{\underline{\alpha}}{}^{\underline{\beta}} = \tfrac{1}{10!\sqrt{-g}} \epsilon^{i_1 \cdots i_{10}} \Pi_{i_1}{}^{a_1} \cdots \Pi_{i_{10}}{}^{a_{10}} (\sigma_{a_{10} \cdots a_1})_{\underline{\alpha}}{}^{\underline{\beta}} \ , \tag{4.12}$$

---

[5] We use $\underline{\alpha}, \underline{\beta}, \cdots = 1, \cdots, 32$ for fermionic, while $a, b, \cdots = 0, 1, \cdots, 9$ for bosonic coordinates.

with $\delta_\kappa E^A \equiv (\delta_\kappa Z^M) E_M{}^A$, and $\Gamma^2 \equiv I$ [29]. The BFFC background constraints are given by (2.10) in [29], satisfying the Bianchi identities (2.1) - (2.6) in [29][6]. These are all consistent with the fermionic $\kappa$-invariance of the action (4.11), in particular with the field strength [29]

$$H_{A_1 \cdots A_{10}} \equiv 10 \nabla_{[A_1} C_{A_2 \cdots A_{10})} + M K_{A_1 \cdots A_{10}} \ . \tag{4.13}$$

which is nothing but the superspace generalization of (4.3).

This feature seems to be common to arbitrary supergravity theories, *i.e.*, the dimensionality of super world-volume $D$ of a $p = (D-1)$-brane coinciding with that of the target space-time $D$ [29]. This gives an independent justification of the generalized Chern-Simons term $MK$ into the field strength $H$, in particular, consistently with D-branes [32] possibly at quantum levels.

As a matter of fact, according to past experiences in supergravity and superstring theories, important geometrical features in superspace at classical level are well-preserved and consistent at quantum levels, unless there is certain anomaly. Since we have the D-brane action supporting the consistency of our backgrounds with our $MK$-type Chern-Simons term, we have a strong foundation to believe the quantum consistency of our mechanism.

## 5. Concluding Remarks

In this Letter, we have presented a new mechanism based on Hodge duality with a Chern-simons term yielding a zero cosmological constant. The key prescription is summarized as

(1) Regard $m$ in the cosmological constant $\Lambda \equiv a m^2/2$ as a '0-form field strength', replacing $m$ by an $x$-dependent scalar field $M(x)$.

(2) Define the $D$-form field strength $H = DdC + MK$ with the Chern-Simons term $MK$.

(3) Add a constraint lagrangian $\mathcal{L}_c \equiv -(a/D!)\epsilon^{[D]} M H_{[D]}$.

(4) The $C$-field equation yields $M \doteq$ const.

(5) The $K$-field equation yields $M \doteq 0 \implies \Lambda = 0$, *i.e.*, the zero cosmological constant.

(6) The $M$-field equation yields the duality $M \doteq (1/D!) e^{-1} \epsilon^{[D]} H_{[D]}$. The difference from the usual duality transformation [22] from 0- to $D$-forms [28] is that both sides vanish due to $M \doteq 0$.

---

[6]The superfields $M$, $N$, $C$ and $H$ in [29] correspond respectively to $C$, $H$, $K$ and $L$ in this present paper.

We have also seen that our mechanism is applicable also to supergravity theories in diverse dimensions [21] consistently with local supersymmetry. As explicit applications, we have seen the examples of 4D supergravity [5] and 10D massive type IIA supergravity [31][27]. Interestingly, the action invariances under supersymmetry allow $^\forall \Lambda \equiv aM^2$, if there is no extra term $\approx MK$ in the field strength $H$. It is the $K$-field equation that forces $M$ to vanish, consistently with local supersymmetry. The vanishing cosmological constant is required by the dynamics of our system, but is not 'artificially' installed by hand.

Even though the consistency with local supersymmetry looks remarkable, it is a natural consequence. In fact, the introduction of the $M$-field [28] was originally motivated by massive type IIA supergravity in 10D [27], and in particular, the maximal-rank field $K$ was introduced in superspace [29]. It is our present mechanism with the particular lagrangian that relates this Chern-Simons term to the vanishing cosmological constant as dynamical $K$-field equation in component formulation. This is because the equation $M \doteq 0$ is a dynamical equation which is always consistent with supersymmetry, superspace or fermionic $\kappa$-symmetry, but is produced only by a component lagrangian term such as $M \wedge H$-term in (4.1).

Our mechanism is based on the non-physical fields $M$, $C$, $K$, which do not affect any physics other than the cosmological constant $\Lambda$ itself. In other words, the exclusion of $\Lambda$ is naturally accomplished by such non-physical fields. Note also that the introduction of the Chern-Simons term $MK$ in the maximal-rank field strength $H = DdC + MK$ inspired by superspace formulation [29] is the key ingredient that has not been presented in component formulation in the past, not to mention the context of cosmological constant problem.

Even though there might be many other mechanisms yielding the same effect $M \doteq 0$, such as using only a scalar density field, those formulations lack underlying principles or symmetries, as our 0-form $D$-form Hodge duality that necessitates our mechanism with the new Chern-Simons $MK$-term in the field strength $H$.

Even though we have discussed only classical systems, there are a few reasons to believe that our mechanism works in quantized systems, at least for supersymmetric cases with fundamental D-brane actions [28]. The first question may be the consistency of duality transformation [22] at quantum levels. This is not too difficult to answer, because the original duality [22] is applied to massive type IIA D-brane formulation [28] for the R-R higher-rank fields of Type IIA superstring. In particular, the existence of 9-form potential is not artificial manipulation, but is well-founded by super-eightbrane formulation [28] which is supposed to be consistent at the quantum level.

A subtler question is the justification of the extra Chern-Simons term $MK$ in the field strength $H$. As is shown in the last section, supporting evidence is that the maximal $D$-rank superfield is related to $p = (D-1)$-brane action [29] compatible with general D-brane formulations [32] and $\kappa$-invariance. Therefore it is natural to expect that our $MK$-type Chern-Simons term in $H$ is respected at each level of quantum loop computations. In fact, according to our past experiences in supergravity or superstring, any geometrical feature in classical superspace backgrounds for extended objects is supposed to be maintained at quantum levels, unless there is anomaly to upset it. Since our $MK$-type Chern-Simons form has a geometrical foundation, we have a good reason to believe its quantum consistency of our mechanism for supersymmetric cases with D-brane actions. Of course, however, we lose such grounds for non-supersymmetric systems with no D-brane action available. For this reason, quantum consistency may fail for non-supersymmetric cases with no fundamental D-brane actions.[7]

Compared with unimodular gravity [14] or unimodular supergravity [15], our mechanism is superior. First, in the former [14][15], the cosmological constant is regarded as an 'initial condition' instead of a parameter fixed by hand. In our formulation, the value $\Lambda = 0$ comes out as the unique solutions to $K$-field equations. In other words, the dynamics of our system determines the special value $M \doteq 0$ out of $^\forall M$ that are allowed for action invariance. Second, our 0-form $D$-form Hodge duality relationships such as (1.2) are motivated mathematically, instead of the particular constraint $\det(e_\mu{}^m) = 1$ which looks rather *ad hoc* and artificial.

The fact that our formulation is successful both for supersymmetric and non-supersymmetric systems strongly suggests the validity of our mechanism even after supersymmetry breakings, at least at classical levels. This is one of the most important features of our mechanism. We re-emphasize that the examples given in this paper cover only a small subset of all the systems that our mechanism can be applied to. By following the simple steps (1) through (6) at the beginning of this section, one can immediately apply our mechanism to gravity or supergravity theories in diverse dimensions. Our mechanism may well be consistent also at quantum levels for supersymmetric backgrounds, if consistent D-brane actions are available.

We are grateful to W. Siegel for important discussions. This work is supported in part by NSF Grant # 0308246.

---

[7]If $\Lambda$ contains also quantum effects, our mechanism in section 2 may still be valid at quantum levels. However, we do not get into the details of this kind for non-supersymmetric systems in this paper, due to the more basic question with non-renormalizability.

## References

[1] *For classic and recent reviews, see e.g.,* S. Weinberg, Rev. Mod. Phys. **61** (1989) 1; E. Witten, *'The Cosmological Constant from the Viewpoint of String Theory'*, Lecture at DM2000, hep-ph/0002297; S.M. Carroll, Living Rev. Rel. **4** (2001) 1, astro-ph/0004075.

[2] M. Green, J.H. Schwarz and E. Witten, *'Superstring Theory'*, Vols. **I** and **II**, Cambridge University Press (1987).

[3] C. Hull and P.K. Townsend, Nucl. Phys. **B438** (1995) 109, hep-th/9410167; E. Witten, Nucl. Phys. **B443** (1995) 85, hep-th/9503124; P.K. Townsend, *'M-theory from its Superalgebra'*, hep-th/9712004; T. Banks, W. Fischler, S.H. Shenker and L. Susskind, Phys. Rev. **D55** (1997) 5112, hep-th/9610043.

[4] P. van Nieuwenhuizen, Phys. Rep. **68C** (1981) 189.

[5] S. Deser and B. Zumino, Phys. Rev. Lett. **38** (1977) 1433.

[6] E. Cremmer, S. Ferrara, L. Girardello and A. van Proeyen, Nucl. Phys. **B212** (1983) 413.

[7] E. Witten, Phys. Lett. **155B** (1985) 151.

[8] S.W. Hawking, Phys. Lett. **195B** (1987) 337; Phys. Rev. **D37** (1988) 904; G. Lavrelashvili, V. Rubakov and P. Tinyakov, JETP Lett. **46** (1987) 167; S. Giddings and A. Strominger, Nucl. Phys. **B306** (1988) 867; *ibid.* **B307** (1988) 854; S. Coleman, Nucl. Phys. **B307** (1988) 867.

[9] T. Banks, I. Klebanov and L. Susskind, Nucl. Phys. **B317** (1989) 665; I. Klebanov and L. Susskind, *'DPF Conf. 1988'*, Talk at Div. of Particles and Fields of APS, Storrs, CT, 1988, p. 786.

[10] A. Einstein, Sizungsber, d. Preuss, Akad. d. Wissensch., Pt. 1, 433 (1919), English translation: H.A. Lorentz, A. Einstein *et al.*, *'The Principle of Relativity'*, Dover, NY (1952).

[11] A.G. Riess *et al.*, Astron. J. **116** (1998) 1009; P.M. Garnavich *et al.*, Astrophys. J. **509** (1998) 74; S. Perlmutter *et al.*, Astrophys. J. **517** (1999) 565.

[12] G.L. Kane, M.J. Perry, A.N. Zytkow, New Astron. **7** (2002) 45, astro-ph/0001197

[13] P.J.E. Peebles and B. Ratra, Astrophys. J. **325** (1988) L17; B. Ratra and P.J.E. Peebles: Phys. Rev. **D37** (1988) 3406; C. Wetterich, Nucl. Phys. **B302** (1988) 668.

[14] J. Anderson, D. Finkelstein, Amer. J. Phys. **39** (8) (1971) 901; J.J. van der Bij, H. van Dam, Y.J. Ng, Physica **A116** (1982) 307; F. Wilczek, Phys. Rev. **D104** (1984) 111; A. Zee, in S.L. Mintz and A. Perlmutter (Eds.), *'Proceedings of the Twentieth Annual Orbis Scientiae on High Energy Physics'*, Plenum, NY (1985); W. Buchmüller and N. Dragon, Phys. Lett. **207B** (1988) 292; W.G. Unruh, Phys. Rev. **D40** (1989) 1048; W.G. Unruh and R.M. Wald, Phys. Rev. **D40** (1989) 2598; M. Henneux and C. Teitelboim, Phys. Lett. **222B** (1989) 481.

[15] H. Nishino and S. Rajpoot, hep-th/0107202, Phys. Lett. **528B** (2002) 259.

[16] *See, e.g.,* C. Vafa and E. Witten, Nucl. Phys. **B431** (1994) 3, hep-th/9408074; *ibid.* **B447** (1995) 261, hep-th/9505053; E. Alvarez, L. Alvarez-Gaume and Y. Lozano,

Nucl. Phys. Proc. Suppl. **41** (1995) 1, `hep-th/9410237`; J.H. Schwarz, Nucl. Phys. Proc. Suppl. **55B** (1997) 1, `hep-th/9607201`; A. Giveon, M. Porrati and E. Rabinovici, Phys. Rep. **244C** (1994) 77, `hep-th/9401139`; N.A. Obers, B. Pioline, Phys. Rep. **318C** (1999) 113, `hep-th/9809039`; S. Forste, Fort. Phys. **50** (2002) 221, `hep-th/0110055`; *and references therein.*

[17] E. Witten, Nucl. Phys. **B443** (1995) 85, `hep-th/9503124`.

[18] E. Witten, Int. Jour. Mod. Phys. **A10** (1995) 1247, `hep-th/9409111`; Mod. Phys. Lett. **A10** (1995) 2153, `hep-th/9506101`.

[19] K. Becker, M. Becker and A. Strominger, Phys. Rev. **D51** (1995) 6603; H. Nishino, Phys. Lett. **370B** (1996) 65.

[20] J. Gates, Jr. and H. Nishino, Phys. Lett. **157B** (1985) 157; *ibid.* **173B** (1986) 46 & 52; Nucl. Phys. **B282** (1987) 1; *ibid.* **B291** (1987) 205; H. Nishino and S.J. Gates, Jr., Phys. Lett. **189B** (1987) 45; H. Nishino, Phys. Lett. **258B** (1991) 104.

[21] *'Supergravity in Diverse Dimensions'*, eds. A. Salam and E. Sezgin (North-Holland/World-Scientific, 1989).

[22] H. Nicolai and P.K. Townsend, Phys. Lett. **98B** (1981) 257.

[23] P.G.O. Freund and M.A. Rubin, Phys. Lett. **97B** (1980) 233.

[24] E. Cremmer, B. Julia and J. Scherk, Phys. Lett. **76B** (1978) 409.

[25] E. Cremmer and B. Julia, Nucl. Phys. **B159** (1979) 141.

[26] A. Aurilia, H. Nicolai and P.K. Townsend, Nucl. Phys. **B176** (1980) 509.

[27] L. Romans, Phys. Lett. **169B** (1986) 374.

[28] E. Bergshoeff, M. de Roo, M. Green, G. Papadopoulos and P.K. Townsend, Nucl. Phys. **B470** (1996) 113, `hep-th/9601150`.

[29] H. Nishino, `hep-th/9901027`, Phys. Lett. **457B** (1999) 51.

[30] H. Nishino and S. Rajpoot, *in preparation.*

[31] F. Gianni and M. Pernici, Phys. Rev. **D30** (1984) 325; M. Huq and M. Namazie, Class. and Quant. Gr. **2** (1985) 293; C. Campbell and P. West, Nucl. Phys. **B243** (1984) 112.

[32] P.K. Townsend, *'P-Brane Democracy'*, in the proceedings of March 1995 PASCOS/Johns Hopkins Conference, `hep-th/9507048`; J. Polchinski, Phys. Rev. Lett. **75** (1995) 184; J. Dai, R.G. Leigh and J. Polchinski, Mod. Phys. Lett. **A4** (1989) 2073.